\documentclass[proceedings,opts]{JHEP}
\conference{TMR conference, Paris, 1999}
\title{String-generated quartic scalar interactions\thanks{Talk delivered by R. Marotta}}
\author{R. Marotta$^{(a)}$ and F. Pezzella$^{(b)}$\\$^{(a)}$ NORDITA, Blegdamsvej 17, DK-2100 Copenhagen, Denmark
\\$^{(b)}$ I.N.F.N. and Universit\`a di Napoli, Compl. Univ. di Monte S. Angelo, I-80126 Napoli, Italy}
\abstract{The {\em cutting and sewing} procedure is used for getting two-loop 
order Feynman diagrams of $\Phi^{4}$-theory with an internal $SU(N)$ symmetry 
group, starting from tachyon amplitudes of the open bosonic string theory. 
In a suitably defined field theory limit, we reproduce the field theory
amplitudes
properly normalized and expressed in the Schwinger parametrization.}
\usepackage{epsfig}
\newcommand{\beq}{\begin{equation}}
\newcommand{\eeq}{\end{equation}}
\newcommand{\beqa}{\begin{eqnarray}}
\newcommand{\eeqa}{\end{eqnarray}}
\newcommand{\nn}{\nonumber}
\newcommand{\eq}[1]{(\ref{#1})}
\newcommand{\ket}[1]{\vert\,{#1}\,\rangle}
\newcommand{\bra}[1]{\langle\,{#1}\,\vert}
\newcommand{\alimit}{\stackrel{\alpha' \rightarrow 0}{\longrightarrow}}
\newcommand{\ra}{\rightarrow}
\newcommand{\zalmu}{z_{\alpha_{\mu}}}
\newcommand{\zbemu}{z_{\beta_{\mu}}}
\newcommand{\zai}{z_{\alpha_i}}
\newcommand{\zbi}{z_{\beta_i}}
\newcommand{\zaj}{z_{\alpha_j}}
\newcommand{\zbj}{z_{\beta_j}}
\newcommand{\imtau}[1]{(2\pi\mbox{Im}\tau)^{-1}_{#1}}
\newcommand{\NP}[1]{ {\it Nucl.~Phys.} {\bf #1}}
\newcommand{\PL}[1]{ {\it Phys.~Lett.} {\bf #1}}
\newcommand{\Prep}[1]{ {\it Phys.~Rep.} {\bf #1}}
\newcommand{\PR}[1]{ {\it Phys.~Rev.} {\bf #1}}
\newcommand{\PRL}[1]{ {\it Phys.~Rev.~Lett.} {\bf #1}}
\newcommand{\PTP}[1]{ {\it Prog.~Theor.~Phys.} {\bf #1}}
\newcommand{\PTPS}[1]{ {\it Prog.~Theor.~Phys.~Suppl.} {\bf #1}}
\newcommand{\MPL}[1]{ {\it Mod.~Phys.~Lett.} {\bf #1}}
\newcommand{\IJMP}[1]{ {\it Int.~Jour.~Mod.~Phys.} {\bf #1}}
\newcommand{\JP}[1]{ {\it J.~Phys.} {\bf #1}:\  Math.~Gen.~}
\renewcommand{\theequation}{\thesection.\arabic{equation}}
\renewcommand{\thefootnote}{\arabic{footnote}}

\begin{document}

It is well-known that a string theory can provide a consistent quantum theory
of gravity, unified with non-abelian gauge theories in the 
so-called {\em zero-slope limit} where
the inverse string tension $\alpha' \rightarrow 0$. The reasons of such a
consistency lie in the fact that this latter is a
physical dimensional parameter acting as an ultraviolet cutoff in the
integrals over loop momenta. Therefore it makes multiloop amplitudes free from 
ultraviolet divergences.

Consequently, string theory provides an alternative technique of computing
field theory amplitudes. In fact, since its scattering amplitudes are 
organized in a very compact form, one can compute, for instance, non-abelian
gauge theory amplitudes by starting from a string theory and performing
the zero-slope limit, rather than using traditional field theory techniques.
The expression of string amplitudes is known
explicitly, including also the measure of integration on 
moduli space, in the case of the bosonic open string for any
perturbative order \cite{D}. 

These interesting features of string theory have led some authors
to use it as an efficient tool to compute gluon amplitudes in Yang-Mills
theory \cite{BK1} $\div$ \cite{DMLRM} or graviton amplitudes in 
quantum gravity \cite{Bgr}, and some improvements have been achieved in 
understanding
the perturbative relations between gravity and gauge theory \cite{BDDPR}. 
Such string-methods have also inspired
some authors in developing very interesting techniques based on the world-line
path-integrals \cite{kajsch}. 

In order to derive field theory amplitudes
from the corresponding string ones one has to start, for example in the case 
of Yang-Mills amplitudes, 
from a given multiloop gluon string amplitude 
and to single out different regions of the moduli space that, in the 
low-energy limit, reproduce different field theory diagrams. This
program has been carried out at one-loop \cite{BDK1} \cite{DMLMR} and, 
at this order, the five-gluon amplitude has been obtained for the 
first time \cite{BDK2}.

This procedure, when extended to the two-loop case
\cite{DMLMR} \cite{MR} \cite{BRY}, becomes difficult to handle 
in the case of amplitudes with external states.
The computational difficulties associated with this kind of 
multiloop Yang-Mills amplitudes, which are inessential for understanding
the field theory limit, can be avoided if amplitudes involving scalar 
particles are considered. In fact, differently from what happens in gauge theories, in string theory
there is not a big conceptual difference between gluon and scalar diagrams.
Therefore one can get more easily from scalar amplitudes
the whole information about the corners of the moduli space reproducing the known field theoretical results:
these regions are exactly the ones giving the correct field theory diagrams
also in the case of gluon amplitudes.

Scalar amplitudes are the ones involving tachyons of the bosonic string 
theory. So one can consider 
a slightly different zero-slope limit of the bosonic
string in which only the lowest tachyonic state, with a mass satisfying
$m^{2}= -1/\alpha'$ is kept. In the case of tree and one-loop diagrams,
this procedure is equivalent to take the zero-slope limit of an old
pre-string dual model with an arbitrary value of the intercept
of the Regge trajectory. It was recognized the inconsistency of this model,
but the field theory limit of tree and one-loop diagrams of this pre-string
dual model was shown to lead to the Feynman diagrams of $\Phi^{3}$ 
theory \cite{S}.

In Ref. \cite{DMLMR}, it has been explicitly shown that
by performing the zero-slope limit as above explained,
one correctly reproduces the Feynman diagrams of $\Phi^{3}$ theory, up
to two-loop order.  

The aim of this talk is to illustrate how the conceptual scheme pursued in 
Ref. \cite{DMLMR} can be extended 
to two-loop amplitudes containing quartic scalar interactions. 
This means that, starting from string amplitudes involving tachyons, 
we perform on them the 
field theory limit in which their mass is fixed and correctly identify 
the corners associated to the different field theory diagrams
of $\Phi^{4}$ theory. Such a program is carried out by pursuing
the so-called {\em sewing and cutting} procedure.

This work is based on the results contained in Ref. \cite{MP2}. We would
like to cite here the article \cite{FMR} where scalar diagrams are obtained 
from string amplitudes through an alternative procedure based on the Schottky
group properties \cite{DPFHLS}.

The plan of the work is the following.

Firstly we illustrate the {\em sewing and cutting} procedure applying it to 
the four tachyon tree amplitude. We define a proper field theory limit where 
quartic scalar interactions are reproduced. Then we check the validity of this
procedure by deriving from the two- and four-tachyon amplitudes at one-loop,
respectively, the {\em tadpole} and the {\em candy} diagram of $\Phi^4$ theory. 
Finally we apply it to the {\em double-candy} diagram.

\section{Tree scalar diagrams from strings}

The starting point is the planar tree scattering amplitude of four on-shell
bosonic open string tachyons with momenta $p_{1}, \dots, p_{4}$ each 
satisfying the mass-shell condition $p^{2}=-m^{2}=\frac{1}{\alpha'}$: 
\[
A_{4}^{(0)}(p_{1},p_2,p_3, p_{4}) = \mbox{Tr} \left[ \lambda^{a_{1}} 
\lambda^{a_{2}} \lambda^{a_{3}}
\lambda^{a_{4}} \right] C_{0} {\cal N}_{t}^{4}
\]
\beq
\times \int_{0}^{1} dz (1-z)^{2 \alpha' p_{2} \cdot p_{3}} z^{2 \alpha' p_{3} \cdot 
p_{4}} 
                                                  \label{A4}
\eeq
where the Koba-Nielsen variables relative to the tachyons labelled by $1,2,4$
have been respectively fixed at $ +\infty,1$ and $0$.

According to the corner of moduli space where the low-energy limit of the amplitude (\ref{A4}) is performed, one can recover, for instance, $\Phi^{3}$- or 
$\Phi^{4}$-scalar diagrams.

In order to understand which regions in moduli space lead
to the different field theory diagrams, one can use the so-called
{\em sewing} {\em and} {\em cutting} procedure. This consists in
starting from a string diagram
and in cutting it in three-point
vertices; next we fix the legs of each three-point vertex at 
$ + \infty$, $1$ and $0$. Then we reconnect 
the diagram by inserting between two three-point vertices a 
suitable propagator acting as a well specified projective transformation.
This is chosen in such a way that its fixed points are just the 
Koba-Nielsen variables of the two legs that have to be sewn. The geometric
role of the propagator is to identify the local coordinate systems defined
around the punctures to be sewn.

The amplitude in Eq. (\ref{A4}) can be expressed
also in terms of the Green functions ${\cal G}^{(0)}(z_{i}, z_{j})$,
defined on the world-sheet in the following way:
\beq
{\cal G}  ^{(0)}(z_{i},z_{j}) = \mbox{log} \left( z_{i} - z_{j} \right)
                               \label{Gf}
\eeq

A necessary intermediate step for deriving tree scalar $\Phi^{4}$-diagrams
is to generate tree diagrams of $\Phi^{3}$-theory.  
With reference to the four-tachyon
tree string diagram, one can see that it can be obtained by 
sewing two three-point vertices as shown in Fig. (1). 
We sew the leg corresponding to the point $0$ in the vertex at the left
hand in Fig. (1a) to the leg corresponding to the point $+\infty$ in the one at the
right hand through a propagator corresponding to the projective transformation
\beq
S(z)=Az    \label{prop}
\eeq
which has $0$ and $+\infty$ as fixed points and the parameter $A$, with
$0 \leq A \leq 1$, as multiplier. Performing the sewing means,
in this procedure, to transform {\em only} the punctures of the three-point 
vertex at the right hand in Fig. (1a) through (\ref{prop}), hence the puncture 
$z_{3}=1$ transforms into $S(1)=A$ while the other two punctures remain
unchanged.

\EPSFIGURE[t]{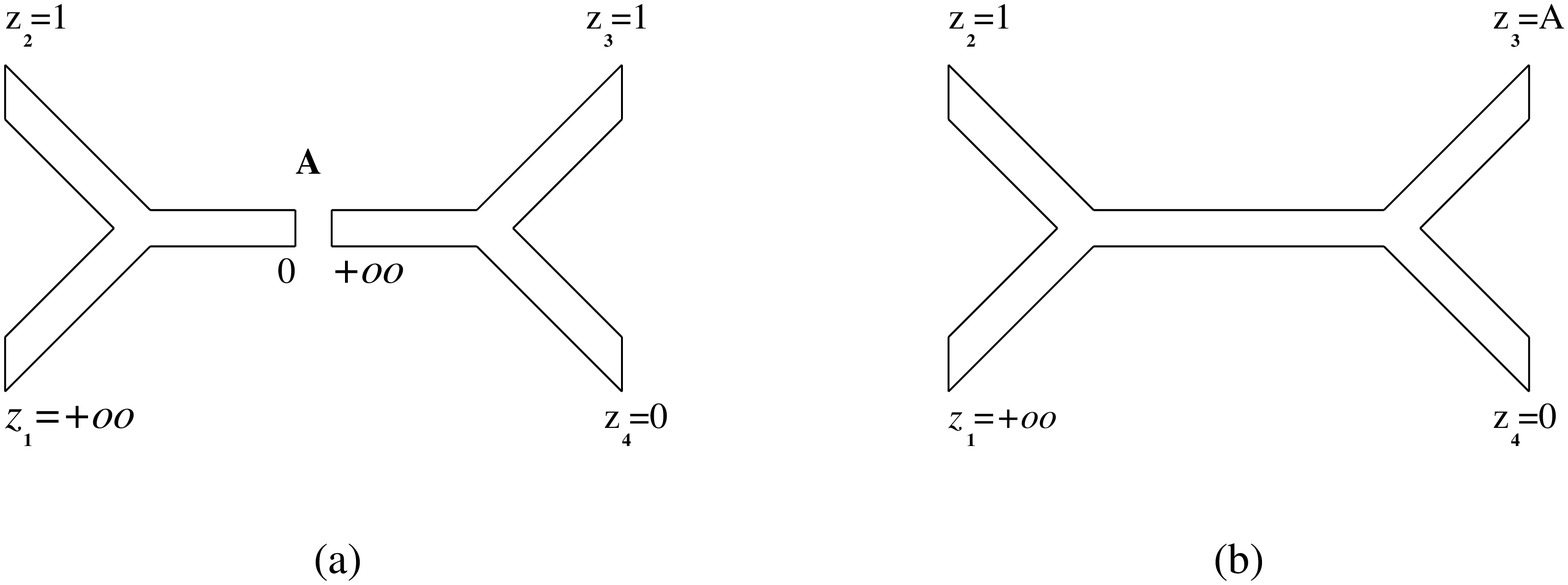,width=12cm}{Sewing of two three vertices}

In general, after the sewing has been 
performed, the Koba-Nielsen variables become
functions of the parameter $A$ appearing in the projective transformation
(\ref{prop}). It is possible to give a simple geometric
interpretation to this parameter, if a correspondence is established
between the projective transformation in Eq. (\ref{prop}) and the 
string propagator, written in terms of the operator 
$e^{-\tau (L_{0}-1)}$. The
latter indeed propagates an open string through
imaginary time $\tau$ and creates a strip of length $\tau$. In fact
the change of variable 
$z=e^{-\tau}$ allows the string propagator to be written as
\[
\frac{1}{L_{0}-1}  =  \int_{0}^{1} dz   z^{L_{0} - 2} 
\]
\beq
 = \int_{0}^{\infty} d \tau \exp \left( -\tau \alpha' \left[ p^{2} + 
\frac{1}{\alpha'} (N-1) \right] \right) \label{LO-1}
\eeq
and to establish the following relation between $\tau$ and $A$: 
\beq
\tau = - \mbox{log}A .   \label{logA}
\eeq  

The multiplier $A$ results to be therefore related to the length of the strip
connecting two three-vertices.

On the other hand, since we want to reproduce tree 
$\Phi^{3}$-theory diagrams
we have to consider a low-energy limit of string amplitudes in which only
tachyons propagate as intermediate states. This is achieved observing
from (\ref{LO-1}) that the only surviving contribution in the limit
$\alpha' \rightarrow 0$ with $\tau \alpha'$ kept fixed is the one coming
from the level $N=0$, i.e. from tachyons with fixed mass given by
$m^{2}= -\frac{1}{\alpha'}$. It is obvious that this also corresponds
to the limit $\tau \rightarrow \infty$ and hence, from (\ref{logA}),
to $ A \rightarrow 0$. From these considerations it seems
natural to introduce the variable $x= \tau \alpha'$ in terms of which
the string propagator (\ref{LO-1}), reproduces, in the above mentioned
limit, the scalar propagator
\[
\frac{1}{p^{2}+m^{2}} = \int_{0}^{\infty} dx e^{-x(p^{2} + m^{2})}
\]
with $x$ being interpreted as the Schwinger proper time.

From a geometrical point of view, one can imagine that the
the strip connecting the two three-vertices, in this field theory limit, 
becomes ``very long and thin'', so that only the lightest states propagate.

By rewriting the amplitude (\ref{A4}) in terms of the Schwinger parameter
$x$ or, equivalently, in terms of the multiplier $A$ we finally get \cite{MP2}:
\beq
A_{4}^{(0)}
=  \frac{1}{8} \mbox{Tr} \left[ \lambda^{a_{1}} \lambda^{a_{2}}
\lambda^{a_{3}} \lambda^{a_{4}} 
\right]
\frac{g^{2}_{\phi^{3}}}{\left[ (p_{1} + p_{2})^{2} + m^{2} \right] } 
                                         \label{A43}
\eeq
where it has been used the well-known relation between $g_{s}$ and 
$g_{\phi^{3}}$ \cite{DMLMR}:
\beq
g_{\phi^{3}} = 16 g_{s} (2 \alpha^{'} )^{ \frac{d-6}{4} } .
\eeq

We are going now to consider a suitable limit of the string four-tachyon
amplitude which can reproduce the diagram
corresponding to the tree four-point vertex of $\Phi^{4}$-theory.
With reference again to the Fig. (1), this diagram has 
to correspond to a limit in which the length of the tube
connecting the two three-vertices composing the string diagram goes to
zero in the limit $\alpha' \rightarrow 0$, i.e.
\[
\tau = \frac{x}{\alpha'}= - \mbox{log} A \rightarrow 0
\]
This corresponds to the limit $A \rightarrow 1$, and hence $z \rightarrow 1$
or, equivalently, $z_{3} \rightarrow z_{2}$. 

In this limit the Green function ${\cal G}^{(0)}(z_{2},z_{3})$ is divergent.
We regularize it by introducing a  cut-off $\epsilon$
on the world-sheet so that
\[
\lim_{z_{2} \rightarrow z_{3}} \mbox{log} \left[ (z_{2}-z_{3})+ \epsilon
\right]  = \mbox{log}
\epsilon
\]
and
\[
\lim_{\alpha' \rightarrow 0} \alpha' \mbox{log} \epsilon = 0
\]
We consider therefore the amplitude $A_{4}^{(0)}$ in Eq. 
(\ref{A4}) in the field theory limit
defined by: 
\[
A=z=1-\epsilon,
\]
\beqa
\alpha' \rightarrow 0 & \mbox{and} & x =  -\alpha' \mbox{log} 
\epsilon \rightarrow 0.   \label{lim}
\eeqa
    
in which it reduces to
\beq
2^{4} \mbox{Tr} \left[ \lambda^{a_{1}} 
\lambda^{a_{2}} \lambda^{a_{3}}
\lambda^{a_{4}} \right] g_{s}^{2} (2 \alpha')^{\frac{d-4}{2}} \label{A44}.
\eeq
The complete amplitude is obtained by performing the sum over 
non cyclic permutations, finally getting a result coincident with the color
ordered vertex generated by the following scalar field theory:

%
%
\beq
{\cal L} = \mbox{Tr} \left[ \partial^{\mu} \phi \partial_{\mu} \phi + m^{2} \phi^2
- \frac{g_{\phi^4}}{4!} \phi^4 \right] \label{phi4}
\eeq
obtaining the {\em matching condition}
\beq
g_{\phi^4} = 4 g_{s} (2 \alpha')^{d/2 - 2} . \label{matching}
\eeq

\section{One-loop $\Phi^{4}$-diagrams from strings}
\setcounter{equation}{0}
\subsection{Tadople diagram}
\EPSFIGURE[t]{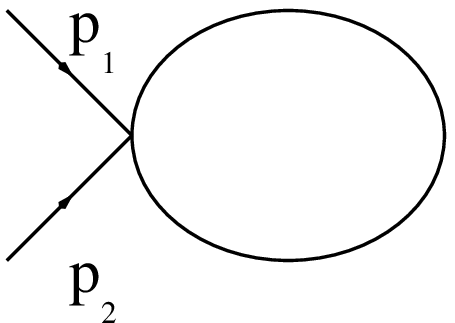,width=3cm}{Tadpole}



In this subsection we show how the tadpole diagram in $\Phi^{4}$-theory can
be derived from string theory.
The starting point will be, this time, the color ordered two-tachyon planar 
amplitude at one-loop: 
\[
A_{2} (p_{1},p_{2}) = N \mbox{Tr} \left[ \lambda^{a_{1}}\lambda^{a_{2}} 
\right] C_{1} \left[ 2 g_{s} ( 2 \alpha' )^{\frac{d-2}{4}} \right]^{2} 
\]
\[
\times \int_{0}^{1} \frac{dk}{k^{2}} \left[ - \frac{1}{2 \pi} \mbox{log} k 
\right]^{-\frac{d}{2}} \prod_{n=1}^{\infty} \left( 1 -k^{n} \right)^{2-d}
\]
\beq
\times \int_{k}^{1} \frac{dz}{z}  \left[ 
\frac{ \exp{ {\cal G}  ^{(1)} ( 1, z )}}{ \sqrt{z} } 
\right]^{2 \alpha' p_{1} \cdot p_{2} }. 
\eeq

We would like now to stress that if we want to reproduce diagrams
of scalar field theories we have to ensure that {\it only} tachyon states
propagate in the loops of string amplitudes. In fact this condition is
fulfilled if small values of the multiplier
$k$ are considered: indeed this parameter plays exactly the same role
as the multiplier $A$ in the tree level amplitudes. 
Therefore an expansion in powers of  $k$ is performed
keeping the most divergent terms. In so doing we get
\[
A_{2}^{(1)} (p_{1},p_{2})= \frac{N}{2} \frac{1}{(4 \pi)^{d/2}} 
\frac{1}{(2 \alpha')^{d/2}}
\left[ 2 g_{s } ( 2 \alpha')^{\frac{d-2}{4}} \right]^{2}
\]
\beq 
\times  \int_{0}^{1} \frac{dk}{k^{2}} \left[ - \frac{1}{2} \log k 
\right]^{-\frac{d}{2}} \int_{k}^{1} \frac{dz}{z} e^{2 \alpha'
{\cal G}^{(1)} ( 1, z )}
\eeq
where the Green function, in the limit we are considering, is
\beq
{\cal G}^{(1)} ( z_{1}, z_{2} ) = \log (z_{1}-z_2) - \frac{1}{2}
\log z_1 z_2 + \frac{ \log^{2} z_1/z_{2} }{2 \log k}
    \label{1g}
\eeq

Our aim is to identify the right limit to get the tadpole diagram in
Fig. (2).
 
Starting from two three-vertices, we sew the leg $0$ with the 
leg $+\infty$ according to the Fig. (3).

\EPSFIGURE[t]{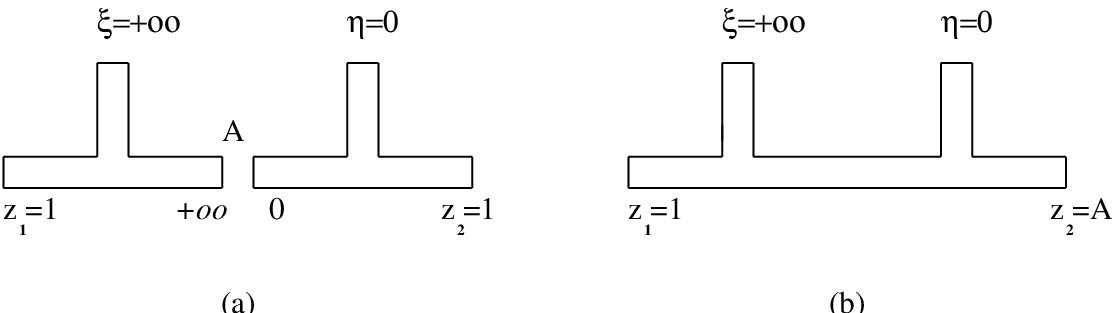,width=12cm}{Sewing for the tadpole diagram}

Such a sewing is performed by considering again the projective transformation
$S(z)=Az$, which has $+\infty$ and $0$ as fixed points and which transforms
$z_{2}= 1$ in the second vertex in Fig. (3a) in the multiplier $A$ getting
the configuration shown in Fig. (3b).

The next step consists in performing a limit in which $z_{2} \rightarrow
z_{1}$, i.e. in which $A \rightarrow 1$ with $\alpha' \mbox{log}
(1-A) \rightarrow 0$, as said before. In this limit we should get 
the tadpole diagram. Indeed we have:

\beq
A_{2}^{(1)}(p_{1},p_{2}) =  \frac{2N}{( 4 \pi)^{d/2}} \frac{1}{2 \alpha'}
g^{2}_{s} \int_{0}^{1} \frac{dk}{k^{2}} \left[ -\frac{1}{2} \log k
\right]^{-d/2}  \label{limA2}
\eeq
By defining:
\[
x = - \alpha' \log k
\]
with $0 \leq x \leq +\infty $, we can rewrite (\ref{limA2}) as follows:
\beq
A_{2}^{(1)}(p_{1},p_{2}) = \frac{ N}{ ( 4 \pi )^{d/2} } \lambda_{\phi^{4}}
\int_{0}^{\infty}dx  e^{-xm^{2}} x^{-d/2} 
\eeq

By using the matching condition established at the tree level (\ref{matching})
we get from string theory the tadpole diagram of $\Phi^4$- theory.

\subsection{Candy diagram}

\EPSFIGURE{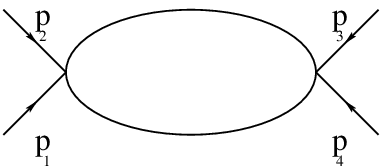,width=6cm}{Candy diagram}


Let us now derive the {\em candy} diagram from
the four-tachyon one-loop amplitude:
\[
A_{4}^{(1)} (p_{1}, p_{2}, p_{3}, p_{4}) =  \frac{N}{( 4 \pi )^{d/2}} 
\mbox{Tr} \left[ \lambda^{a_{1}}
\lambda^{a_{2}} \lambda^{a_{3}} \lambda^{a_{4}} \right]
\]
\[ 
\times \frac{1}{( 2 \alpha^{'} )^{d/2} }
\left[ 2 g_{s} ( 2 \alpha^{'} )^{\frac{d-2}{4}} \right]^{4}
\int_{0}^{1} \frac{dk}{k^{2}} \left[ -\frac{1}{2} \mbox{log} k
\right]^{-\frac{d}{2}} 
\]
\beq
\times \int_{k}^{1} \frac{dz_{4}}{z_{4}}
\int_{z_{4}}^{1} \frac{dz_{3}}{z_3} \int_{z_{3}}^{1} 
\frac{dz_{2}}{z_2} 
\]
\[
\prod_{i<j=1}^{4} \left[ 
\frac{ \exp{ ( {\cal G}(z_{i},z_{j} )) }}{\sqrt{z_i z_j}} 
\right]^{2 \alpha'p_{i} \cdot p_{j}} \label{candy}
\eeq

The diagram relative to this amplitude can be obtained by means of the sewing
procedure illustrated in Fig. (5).

\EPSFIGURE[t]{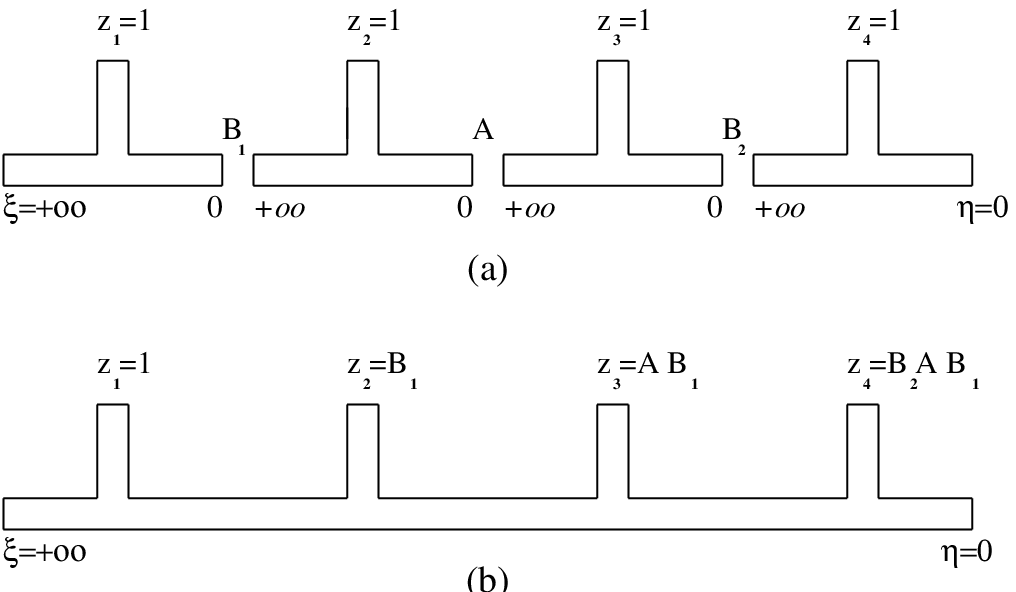,width=8cm}{Sewing for the candy diagram}



The four-particle vertices of the candy diagram can be generated by
the corner of the moduli space where the
Koba-Nielsen variables $z_{1} \rightarrow z_{2}$ and $z_{3} \rightarrow z_{4}$.
This is performed by considering the limit in which the multipliers 
$B_{i}$ $(i=1,2) \rightarrow 1$. We stress here that, in this limit, the Green
functions ${\cal G} (z_{1},z_{2})$ and ${\cal G} (z_{3},z_{4})$ result to
be divergent and we regularize them by introducing
a cut-off $\epsilon$ on the world-sheet so that $B_{i}=1-\epsilon$. In this
limit the length of the strips connecting the three-vertices become very
short and in this way the four-particle vertices of the diagram 
in consideration are generated.
Furthermore, in order to select in the loop only the lightest states, we also
take the limit in which the multiplier $A \rightarrow 0$, and, after 
having performed both the limits, we send the cut-off to zero 
in all the regular expressions.

From these geometrical considerations that shed light on the different roles
played by the multipliers $A$-like and $B$-like, we select the following
corner of the moduli space reproducing the candy diagram of $\Phi^4$-theory:

\beq
A \rightarrow 0 \hspace{1cm} B_{i}=1-\epsilon \rightarrow 1   \label{corner} 
\label{corncan}
\eeq

Let us now evaluate the amplitude (\ref{candy}) in the corner (\ref{corner}).

The first step consists in rewriting, in this region of the moduli space,
the measure and the integration region in the amplitude (\ref{candy}).

The ordering of the Koba-Nielsen variables determines the integration
regions of the multipliers $A$ and $B_{i}$  in terms of which the whole
amplitude is expressed, after the sewing. More precisely, in the limits
(\ref{corncan}), one gets:

\beqa
&&\int_0^1\frac{d\, k}{k^2}
\int_k^1\frac{d\, z_2}{z_2}\int_k^{z_2}\frac{d\, z_3}{z_3}
\int_k^{z_3}\frac{d\, z_4}{z_4}\nn
\\
&&\simeq\int_0^1\frac{d\, k}{k^2}\int_{k}^1\frac{d A}{A}+
\, O \left(k\right)
\eeqa

For this diagram, the proper times associated to the single propagators
in the loop, are identified with the Schwinger parameters

\beq
t_1 = - \alpha' \log k/A \hspace{1cm} t_2 = -\alpha' \log A 
\eeq
where $k$ has to be understood as the proper time of the whole
loop.

The Green functions defined in (\ref{1g}), in this limit, drastically simplify.
 
%
In particular the Green function $2 \alpha' {\cal G}(z_1, z_3)$, when
written in terms of the Schwinger parameters, becomes
\beq
2 \alpha' {\cal G} (z_1, z_3) = t_2 - \frac{{t_2}^{2}}{t_1 + t_2} \,\, .
\eeq

By expressing the full amplitude in terms of $t_{1}$ and $t_{2}$ one gets:
\beqa
&& A_{4} = \frac{N}{(4 \pi)^{d/2}}
\frac{1}{2} d^{a_1 a_2 l}d^{a_3 a_4 l} \left[ 2^{6} g^{4}_{s} (2 \alpha')^{d-4} \right] \nonumber \\
&& \times \int_{0}^{\infty} dt_1 \int_{0}^{\infty} dt_{2} (t_{1} + t_{2} )^{-d/2}
e^{-m^{2} (t_1 + t_2)} \nonumber\\
&& \times  e^{-(p_1 + p_2)^2  \left[
 t_2 - \frac{{t_2}^{2}}{t_1 + t_2}\right]}
\eeqa

Once again we have the right result in field theory by using the
matching condition (\ref{matching}).

\section{A two-loop diagram: double-candy}
\EPSFIGURE[t]{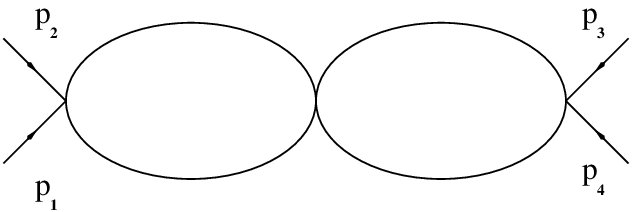,width=6cm}{Double-candy}

In this section we show how to get the {\em double-candy} diagram  of 
$\Phi^4$ theory, Fig. (6), starting from the 
two-loop four-tachyon amplitude in bosonic string theory:
\[
A_{4}^{(2)}(p_1 p_2 p_3 p_4)=N^2 Tr \left[ \lambda^{a_1} \lambda^{a_2}
\lambda^{a_{3}} \lambda^{a_4}
\right]
C_2 N_0^4
\]
\beq
\times
\int \left[dm\right]_2^4 \prod_{i<j}\left[ \frac{ \exp {\cal G}^{(2)}(z_i,\,z_j)}{
\sqrt{ V_i^{'} (0)\, V_j^{'}(0)} } \right]^{2\alpha^{'} p_i\cdot p_j}
\label{2ta}
\eeq
where the expressions for $V^{'}_i(0)$ are given by:
\beq
(V_{i}^{'}(0) )^{-1}=\left| \frac{1}{z_i - \rho_a}-\frac{1}{z_i - \rho_b}\right|
\label{vi}
\eeq
with  $\rho_a$ and $\rho_b$ depending on the position of $z_i$ and being the 
two fixed points on the left and on the right hand of $z_i$.

We expand the previous amplitude  
for small values of the  multipliers $k_{\mu}$ keeping
the most divergent contribution that is the one corresponding to
the tachyon state and again the Green functions reduce to the
following form \cite{DMLMR}:
  
\[
{\cal G}^{(2)}(z_i,\,z_j)= \log(z_i\, -\, z_j)
\]
\beq
+\frac{ \log^2 T \log k_2 +
\log^2 U \log k_1 - 2 \, \log T  \log U  \log S }{
2(\log k_1 \log k_2 - \log^2S)}\nn\\
\label{gf}
\eeq

with

\beqa
S & = &\frac{(\eta_1 -\eta_2)(\xi_1-\xi_2)}{(\xi_1-\eta_2)(\eta_1-\xi_2)} 
\nn\\
T & = & \frac{(z_j -\eta_1)(z_i-\xi_1)}{(z_j-\xi_1)(z_i-\eta_1)} 
\nn \\
U & = & \frac{(z_j -\eta_2)(z_i-\xi_2)}{(z_i-\eta_2)(z_j-\xi_2)} 
\eeqa

The measure, once used the projective invariance to fix $z_4=1$, 
$\xi_2=+\infty$ and $\eta_2=0$, becomes:

\[
\left[ d m\right]_{2}^{4}=\prod_{i=1}^{3}
\frac{d z_i}{V_i^{'}(0)}
\prod_{\mu=1}^{2} \frac{d k_{\mu}}{k_{\mu}^{2}} \frac{d \xi_1 d\eta_1}{
(\xi_1 -\eta_1)^2}
\]
\beq
\times \left[det \left( -i \tau_{\mu\nu}\right)\right]^{-d/2}  
\label{2ms}
\eeq
where $\tau_{\mu\nu}$ is the period matrix in the limit of small multipliers
\cite{MP2}.

Let us now identify the corner of the moduli space that, in the field theory 
limit, reproduces the two-loop candy diagram, according to our procedure.
In Fig. (8) it is shown the final configuration that we reach 
applying the sewing procedure with the following projective 
transformations:    

\beq
\begin{array}{ccc}
S_i= B_i \, z &  \hat{S}_1=A_1\, z  & \hat{S}_2=A_2\, z
\end{array}
\label{prt}
\eeq
with $i = 1,2,3$.

Once the sewing procedure is completed, the Koba-Nielsen variables 
and the moduli of the surface, are expressed in terms of the multiplier of the
transformations according the correspondence shown in Fig. (8).

If we want to obtain 
the four-particle vertices peculiar of the two-loop candy diagram,
we have to take in consideration the corner of the Koba-Nielsen 
variables characterized by $z_1\rightarrow z_2$, 
$z_3\rightarrow z_4$ and by the modulo
$\xi_1\rightarrow 1$. This configuration is achieved considering 
the limits in which 
$B_i\rightarrow 1$ and introducing the suitable regularizators, when necessary.

Furthermore, considering also the limit $A_i\rightarrow 0$, we select scalar
particles in the other internal legs.

The corner of  moduli space reproducing the $\Phi^4$ scalar diagram 
illustrated in Fig.(6) is

\beq
\begin{array}{lr}
A_{i}\rightarrow 0  & \hspace*{1cm}B_{i}= 1 -\epsilon  \,\,\, .  
\end{array}
\label{cor}
\eeq

Let us now evaluate  the amplitude (\ref{2ta}) in this corner.

The Green functions (\ref{gf}) are then evaluated in the limit (\ref{cor})
where they take a simple form and the same is done for the local coordinates $V_i^{'}(0)$
and for the measure.




As regards the integration region, we observe that the sewing procedure 
determines an ordering of the Koba-Nielsen variables and of the fixed points as
shown in Fig. (8).

%

In the field theory limit (\ref{cor}) we integrate the multipliers 
$B$-like between $0$ and $1$ and the multipliers $A$-like, between $0$ and
$\delta$ being $\delta$, a positive infinitesimal quantity.

The Schwinger parameters in this case 
are related to the $A_i$'s 
by the following 
relations \cite{DMLMR}:

\beq
t_{i+2}=-\alpha^{`} \log A_i,
t_1=-\alpha^{'} \log \frac{k_1}{A_2},
t_2= -\alpha^{'} \log \frac{k_2}{A_1} 
\label{sch}
\eeq
with $i=1,2$.

Rewriting the Green functions and the measure 
in terms of the Schwinger parameters we get:

%

\beqa
&&A_{4}^{(2)}(p_1\cdots p_4)=\frac{N^2}{(4\pi)^2} d^{a_1\,a_2\,l} \,
d^{a_3\,a_4\,l} \nn \\
&&\times \frac{ \left[ 2^4 g_s^2 (2\alpha^{'})^{\frac{d-4}{2}}\right]^3}{2^5}
\int_{0}^{\infty} \prod_{i=1}^4 d t_i e^{  -m^2(t_1+t_2+t_3+t_4)} \nn\\
&& \times
(t_1+t_4)^{-d/2} (t_2+t_3)^{-d/2} \nn\\
&& \times e^{-(p_1+p_2)^2 \left[ 
\frac{t_1\, t_4}{t_1\, +\, t_4} + \frac{t_2 \, t_3}{t_2 + t_3}\right]}
\label{am}
\eeqa
where a sum over inequivalent permutations of the external particles 
has been done analogously as in the one-loop candy-diagram case.

Now using the matching condition (\ref{matching}),
we get the same result, including the overall factor, as the one 
obtained in field theory. 

In conclusion, we have used the {\em sewing and cutting} procedure in order to show
how $\Phi^4$-theory diagrams can be reproduced from string amplitudes,
up to two loop-order. The whole information so obtained can be in principle
extendible to Yang-Mills diagrams involving quartic interactions.

\EPSFIGURE[t]{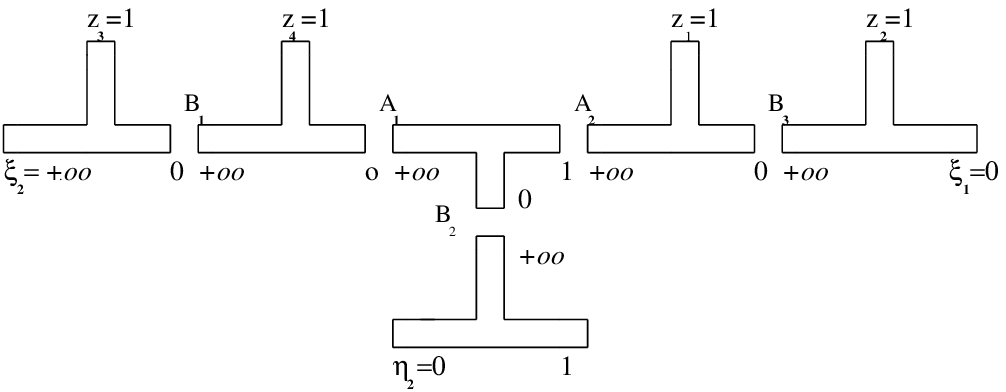,width=10cm}{Sewing of the two-loop candy diagram}


\EPSFIGURE[t]{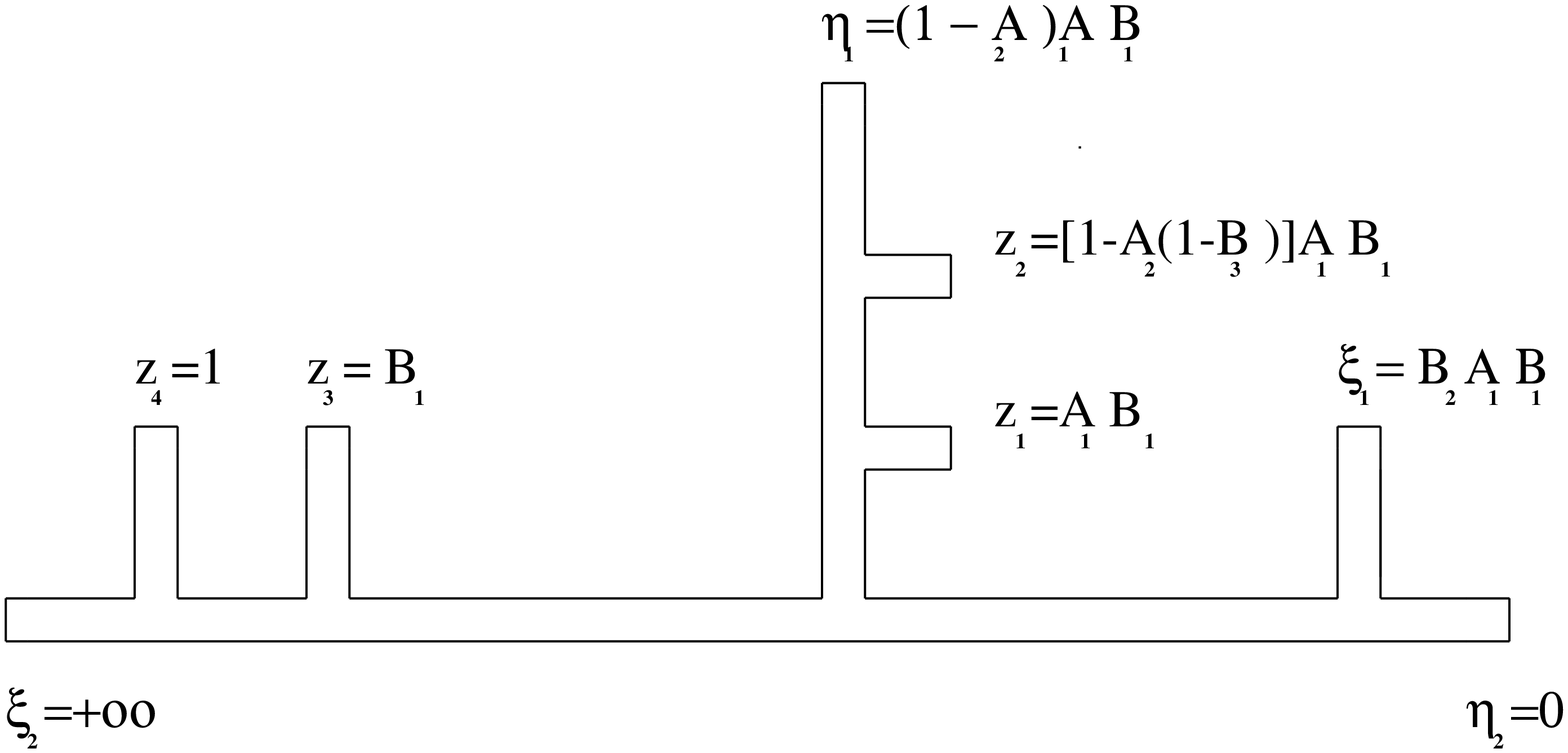,width=10cm}{Sewing configuration for the two-loop 
candy diagram}



\acknowledgments{

We would like to thank P. Di Vecchia, A. Frizzo, L. Magnea, R. Russo and
M.G. Schmidt for helpful discussions.

R.M. and F.P. thank respectively Universit\`a di Napoli and NORDITA 
for their kind hospitality during different stages of their work.}

\end{document}